\documentclass[letterpaper,twocolumn,10pt]{article}
\usepackage{usenix,epsfig,endnotes}
\usepackage{float}
\restylefloat{table}
\usepackage{multirow}

\usepackage{graphicx}
\usepackage{amsmath}
\usepackage{amssymb}
\usepackage{booktabs}
\usepackage{comment}
\usepackage{algorithm}
\usepackage{algpseudocode}
\usepackage{float}
\usepackage{subcaption}

\begin{document}

\date{}

\title{\Large \bf MisGUIDE : Defense Against Data-Free Deep Learning Model Extraction}

\author{
{\rm Mahendra Gurve, Sankar Behera, Satyadev Ahlawat, Yamuna Prasad }\\
Indian Institute of Technology, Jammu.}

\maketitle

\thispagestyle{empty}

\subsection*{Abstract}
The rise of Machine Learning as a Service (MLaaS) has led to the widespread deployment of machine learning models trained on diverse datasets. These models are employed for predictive services through APIs, raising concerns about the security and confidentiality of the models due to emerging vulnerabilities in prediction APIs. Of particular concern are model cloning attacks, where individuals with limited data and no knowledge of the training dataset manage to replicate a victim model's functionality through black-box query access. This commonly entails generating adversarial queries to query the victim model, thereby creating a labeled dataset.

This paper proposes "MisGUIDE", a two-step defense framework for Deep Learning models that disrupts the adversarial sample generation process by providing a probabilistic response when the query is deemed OOD. The first step employs a Vision Transformer-based framework to identify OOD queries, while the second step perturbs the response for such queries, introducing a probabilistic loss function to MisGUIDE the attackers. The aim of the proposed defense method is to reduce the accuracy of the cloned model while maintaining accuracy on authentic queries. Extensive experiments conducted on two benchmark datasets demonstrate that the proposed framework significantly enhances the resistance against state-of-the-art data-free model extraction in black-box settings.

\section{Introduction}
\label{sec:intro}

The landscape of deep learning has undergone a remarkable transformation in recent years, fostering its integration into mission-critical applications such as self-driving automobiles, surveillance systems, and biomedical research, as well as cutting-edge developments like Large Language Models and Generative AI. However, the process of training a high-performing model demands the utilization of high-quality extensive datasets, computational resources with high-performance capabilities, and the knowledge of human experts. Both the collection of data and the training procedure are resource-intensive in terms of cost and time. The significant commercial value of these models has become the status of critical intellectual property for the model owners. This has led to the practice of protecting or securing specific model features, like the model architecture, training algorithms, training data, and hyperparameters, as closely held trade secrets within a black-box framework. The secrecy of their proprietary model is ensured by the black-box models, which generate predictions for user query without revealing any internal information to the service users.

The deployment of an end-to-end infrastructure for utilizing Deep Neural Networks as a service has been adopted as a business model by major MLaaS (Machine Learning as a Service) platform providers including Google, Amazon, Microsoft, OpenAI, Roboflow, and BigML \cite{Chen2023}. These services simplify the deployment of machine learning models, making them more accessible to businesses and developers. In the paradigm of Machine Learning as a Service (MLaaS), service providers receive training data from trusted customers. These providers invest substantial resources in the design and training of models that exhibit high performance. These models are then made available to the public for usage based on a pay-per-query basis. Although, black box Neural Networks within MLaaS hold potential, their monetary worth is compromised by their vulnerability to adversaries that seek to steal the model functionality and properties, as well as bypass the pay-per-query system of the service\cite{Orekondy2020}.

Regrettably, recent studies in \cite{Juuti2019, Papernot2017}, \cite{Tramer2016, Rosenthal2023, Truong2021, Kariyappa2021, Sanyal2022} have exposed the vulnerability of machine learning models to model stealing attacks. Such attacks allow adversaries to generate clone models that, in some instances, reach up to 99\% similarity with the victim model in terms of generalization accuracy. Model extraction, commonly known as model cloning, poses a dual threat. Firstly, it entails financial implications, as competitors could replicate intellectual property, circumventing expenses related to fully trained models and querying MLaaS APIs \cite{Battis2022}. Secondly, model extraction raises more severe concerns like potential violations 
of model privacy through membership inference attacks \cite{Salem2019, Shokri2017} and model inversion attacks \cite{Fredrikson2015, Wang2021}. By generating adversarial examples, it could potentially compromise the integrity of the model.These intricate risks not only damage a business's reputation but also elevate the likelihood of 
facing significant penalties for breaching data protection laws.

Model extraction attacks employ various methods to generate data for querying a victim model which enables adversaries to extract valuable information from the victim model without direct access to its training data. These methods encompass the use of synthetic data \cite{Juuti2019, Papernot2017}, where iterative perturbations are applied to in-distribution seed examples, and surrogate data \cite{Tramer2016}, which involves querying the victim model using a surrogate dataset. More recently, attacks \cite{Rosenthal2023, Truong2021, Kariyappa2021, Sanyal2022} have incorporated data generative methods, leveraging technologies like Generative Adversarial Networks (GANs)\cite{Goodfellow2020} to produce new data samples from the random noise. 

In this work, we primarily focus on model extraction attacks such as DFME \cite{Truong2021} and DISguide\cite{Rosenthal2023} that utilize Generative Adversarial Networks (GANs) to generate query samples. These attacks are commonly referred to as data-free model extraction attacks and the objective of these attacks is to maximize the dissimilarity between the victim model and the clone model, allowing the clone model to enhance its understanding and acquire previously undiscovered knowledge from the queries. However, these methods have two major disadvantages: Firstly, a large number of queries (usually in millions) are required to effectively generate the clone model. Secondly, it has been observed that a substantial proportion of queries generated by GANs fall into the OOD category with respect to the data distribution of the victim model.

Motivated by the limitations in existing attack methods \cite{Truong2021, Rosenthal2023, Sanyal2022}, we introduce MisGUIDE as a defense mechanism against model stealing (or cloning) attacks. The defense mechanism of MisGUIDE relies on the observation that contemporary model extraction attack methods use a generative deep learning model to generate new query samples from a distribution using random noise. These methods generate a significant number of OOD samples to query the victim model. A Vision Transformer based OOD detector is employed to identify 'suspicious' query, which could potentially originate from an adversary in the proposed MisGUIDE framework. The response to the query from victim model is generated based on a probabilistic threshold. MisGUIDE deliberately provides inaccurate predictions for queries identified as OOD based on the probabilistic threshold, while in-distribution (ID) queries receive accurate predictions. This strategy leads to the mislabeling of a significant portion of the adversary's query. Training a model on this mislabeled dataset results in a clone model with poor generalization accuracy. 

The proposed approach aligns with recent defense methods \cite{Orekondy2020, lee2018defending, kariyappa2020defending, tang2024modelguard} that leverage similar insights into misleading the adversary by introducing perturbations to the model's predictions. Related works \cite{Orekondy2020, lee2018defending} introduce a defense mechanism to mitigate the effectiveness of model extraction attacks by perturbing the output probabilities returned by the MLaaS models. Although, these defenses are effective in reducing the damage if the model returns a confidence score vector (soft labels) but not applicable when the model returns only output class labels (hard labels). The adaptive defense mechanism method outlined in \cite{lee2018defending} utilizes an OOD detector to perturb the confidence score of all OOD queries. However, in practical scenarios, there exists a potential risk where in-distribution queries could be erroneously classified as OOD by the OOD detector, resulting in inaccurate output for legitimate users. 

In this paper, we introduce a novel defense mechanism, MisGUIDE, to protect deep learning models against model cloning attacks involving adversarial queries. Our proposed defense framework exhibits scalability and effectiveness, offering robust protection against both soft and hard label-based model extraction attacks, outperforming other state-of-the-art perturbation-based defense strategies. Our contributions can be summarized as follows:
\begin{itemize}
\item 
This work demonstrates the effectiveness of using a pre-trained Vision Transformer as an OOD detector to identify malicious queries generated by adversaries, thereby enhancing protection against model extraction attacks.
\item 
A novel defense framework "MisGUIDE" to safeguard the victim from the model extraction attacks is proposed. It effectively defends data-free model extraction under both soft and hard label settings, without relying on assumptions about attackers' models or data, distinguishing it from existing defenses.
\item 
A probabilistic threshold to provide random predictions for the adversary's OOD queries is introduced. This enables model extraction more challenging while maintaining the victim model's accuracy on benign queries.
\item 
The MisGUIDE framework is evaluated on two benchmark CIFAR10 and CIFAR100 datasets. The experimental results show that MisGUIDE outperforms state-of-the-art defenses by achieving a significant balance between high test accuracy and low cloning accuracy when defending against recent model extraction attacks.
\end{itemize}

The rest of the paper is organised as follows: Section 2 presents the related work and preliminaries on the state-of-the-art attacks and defense mechanism. In Section 3, the proposed MisGUIDE framework is described. The experiments and results are demonstrated in Section 4. The work is concluded alongwith future directions in Section 5.

\section{Related Work and Preliminaries}
\subsection{Model extraction attacks}
Model extraction attacks are a key issue in the field of machine learning security \cite{gao2021tenet}\cite{milli2019model}\cite{Chen2023}\cite{pal2019framework}\cite{battis2022transformer}\cite{correia-silva2018copycat}. \cite{Tramer2016} introduced model extraction attacks on linear ML models using prediction APIs queries. \cite{Papernot2017} propose a method for stealing a black-box neural network model by using a Jacobian-based heuristic to identify examples defining the decision boundary of the target model when the attacker has limited training data.\cite{wang2018stealing} focus on stealing a machine learning parameter, specifically the regularization parameter. \cite{milli2019model} developed a gradient-based approach for two-layer ReLU network extraction.\cite{pal2019framework} used methods of active learning for implementing model extraction attacks on Deep Neural Networks (DNNs) for image and text classification. In another addition, \cite{Orekondy2020} proposed training a attack model to generate query-prediction pairs that beat the target model's accuracy.
However, These approaches face challenges in achieving satisfactory performance without an appropriate surrogate dataset. This limitation has led to the development of data-free techniques, exemplified by recent methods like MAZE \cite{Kariyappa2021} and DFME \cite{Truong2021}. These approaches aim to extract models using synthetic data generated by GANs. Specifically, the generator is trained to produce images maximizing the dissimilarity score between the clone and victim models. While these data-free methods are computationally expensive, demanding a substantial number of queries (approximately 20 million) to the victim model in a black-box setting, they also assume access to the softmax vector from the victim model. In contrast, DISguide\cite{Rosenthal2023} and DFMS\cite{Sanyal2022} introduce data-free model extraction attacks, targeting a practical scenario where the adversary only has access to hard labels from the victim model and no access to the victim's training data. These attacks methods excels in both hard-label and soft-label scenarios, employing a novel GAN framework with training on synthetic data to boost the clone's accuracy. These methods improves the performance of attacks in terms of cloning accuracy and number of queries required. Most of the a standard defense involves providing only the hard label prediction without providing soft label with maintaining accuracy for MLaaS providers. To attack on such defences,   DFMS introduces independent generator training, collaborating with a discriminator for label distribution optimization. In contrast, DisGUIDE operates in a data-free manner without requiring proxy data.

\subsection{Defenses against model extraction attacks}
Preventing model cloning attacks requires establishing a careful balance between denying or disrupting adversarial queries while allowing benign ones. The goal is to prevent adversarial attempts at extraction with the least amount of interference to regular service for authorized users.

Currently, the most common techniques to prevent model cloning involve implementing modifications to the model outputs in order to cause disruption with the adversary's optimization procedure.Lee et al.\cite{lee2018simple} devised a defense mechanism against model stealing attacks by incorporating misleading perturbations into the model's outputs. while preserving top-1 level accuracy. Alternative defense strategies, such as Maximizing Angular Deviation (MAD) \cite{Orekondy2020}, introduce controllable intensity perturbations to the model outputs, safeguarding against model stealing attacks but at the cost of benign accuracy. 
Another defensive method takes advantage on the adversaries' data constraints by prompting the victim model to generate distinct outputs for inputs that fall within the known distribution and those that fall outside of it. Kariyappa et al. \cite{kariyappa2020defending} proposed the Adaptive Misinformation (AM) defense, which detects OOD inputs and alters the outputs in order to mislead adversaries in an adaptive way.

Additional defenses against model stealing include monitoring the information exposed to each user's model, allowing the system to reject responses \cite{kesarwani2018model}. Alternatively, some approaches utilize alert mechanisms \cite{pal2021stateful} to enhance security.Some methods includes the implementation of digital watermarking \cite{adi2018turning}\cite{cao2021ipguard}. Watermarking technique involves embedding a detectable watermark within the targeted model, allowing for the identification of stolen models.

\subsection{Preliminaries}

The target model, denoted as $\mathcal{M}_{v}$, trained on a labeled training dataset $\mathcal{X}_{v}$ and evaluated on test dataset $\mathcal{X}_{t}$. This model is deployed as a black-box system under Machine Learning as a Service (MLaaS) paradigm. The adversary can send query inputs to the target model (potentially through an API) and the resultant output may encompass either the predicted class label (regarded as hard labels) or the predicted softmax score (a probability vector regarded as soft labels).

\subsubsection{Adversary’s goal}
The adversary's main goal is to generate a clone model $\mathcal{M'}_{v}$ that performs similar to the victim model $\mathcal{M}_{v}$. The adversary's objective defined in Equation \eqref{mis_eq1} maximizes the cloning accuracy, which represents the accuracy achieved by the clone model $\mathcal{M'}_{v}$ on the victim's test set $\mathcal{X}_{t}$.
\begin{equation}
\label{mis_eq1}
\max_{\theta'} \mathbb{E}_{x \sim \mathcal{D}(X)} \textbf{Acc[}\mathcal{M'}_{v}(x; \theta')\textbf{]}
\end{equation}
where $\mathcal{D}(X)$ denotes the data distribution of the problem domain on which the target model is trained. Every sample $x$ belongs to $\mathcal{D(X)}$. $\theta'$ represents the parameter set of the cloned model $M'_{v}$. 

\subsubsection{Adversary’s knowledge and capabilities: }

\begin{itemize}
\item 
\textbf{Target model knowledge :}
Most of the cloning attacks are designed under black-box settings. In this black-box setting, the adversary has no access to the parameter set $\theta$ of the victim model $\mathcal{M}_{v}$. The cloning attack is conducted without knowing the information regarding the victim model's architecture, parameters and hyper-parameters. The adversary is only aware of the inference task performed by 
$\mathcal{M}_{v}$ and it can query $\mathcal{M}_{v}$ to obtain either class labels (hard label) or softmax scores (soft label). 
\item 
\textbf{Data knowledge :}
It is considered that the adversary does not know the training data distribution $\mathcal{D}(X)$. The adversary starts the attack by supplying the queries from its own data distribution $\mathcal{D}_{A}(X)$ to the victim model and gradually adapting the victim's training data distribution $\mathcal{D}(X)$ by reducing the loss from victim model. At the start the adversaries data distribution $\mathcal{D}_{A}(X)$ markedly deviates from $\mathcal{D}(X)$. 
The queries made by the adversary is categorized as (OOD) and the queries sampled from $\mathcal{D}(X)$ is categorized as in-distribution (ID).

\item 
\textbf{Query limits: }It is assumed that the adversary possesses unrestricted access to the model $\mathcal{M}_{v}$ via a prediction API, enabling them to execute an unlimited number of queries. The symbol $Q$ denotes the total number of queries made by the adversary.
\end{itemize}

\begin{figure*}[]
  \centering
   \includegraphics[width=0.8\linewidth]{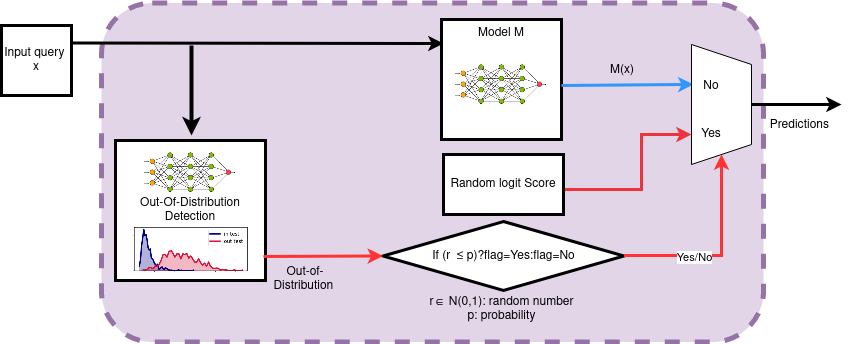}

   \caption{MisGUIDE framework, highlighting its core components: the Victim Model (M), an OOD Detector, a Misguiding Function introducing controlled randomness, and a Switch Mechanism dynamically deciding accurate or intentionally incorrect predictions.}
   \label{fig:Model}
\end{figure*}

\subsubsection{Data Free Model Extraction Attacks (DFME) \cite{Truong2021}}

An adversary can query the defender's model using synthetic or surrogate data when an actual dataset is unavailable. From these pairs of queries and predictions, an adversary can construct a labeled dataset to train the clone model. Generally, these approaches are based on knowledge distillation to train a "student" model (clone model) using predictions from a "teacher" model (defender's model). 

Our focus is on analyzing and defending the attacks where the adversary lacks information about the distribution of the victim's training data and uses the synthetic data to execute model stealing attacks. Recent studies [15, 21] demonstrate the feasibility of utilizing a synthetic data distribution to replicate the functionality of black-box models. DFME aims to determine the clone (student) model's $\mathcal{M'}$ parameters $\theta$ that minimize errors between the clone $\mathcal{M'}(x)$and victim predictions $\mathcal{M}(x)$,where every x $\in$ $\mathcal{D}(X)$.  
\begin{equation}
\underset{\theta}{\mathrm{argmin}}\mathbb{P}_{x \sim \mathcal{D}(X)} (\underset{i}{\mathrm{argmin}}\mathcal{M}_{vi}\neq \underset{i}{\mathrm{argmin}}\mathcal{M'}_{vi})
\end{equation}
. Since the victim's domain is not publicly available, a data-free model extraction attack minimizes the student's error on a synthesized dataset $\mathcal{D}_{A}(X)$. This is achieved by optimizing a loss function 
$\mathcal{L}$
that measures disagreement between the victim and student predictions. 
\begin{equation}
\underset{\theta}{\mathrm{argmin}}\mathbb{E}_{x \sim \mathcal{D}(X)} \left [ \mathcal{L}\left (  \mathcal{M}_{v},\mathcal{M'}_{v}\right ) \right ]
\end{equation}

The performance of DFME \cite{Truong2021}  is improved by DisGUIDE \cite{Rosenthal2023}, which incorporates a joint loss that combines the disagreement loss and class diversity loss during the training of the generator. This directs the generator to generate query samples that induce disagreement among clone models and variation in their predictions. Consequently, the training phase of clones is enhanced in terms of effectiveness and efficiency, enabling DisGUIDE to derive models with cutting-edge accuracy while necessitating a reduced query budget. Equation 1 presents the joint loss function, designated as $\mathcal{L}_{G}$, which is the aggregate of the new disagreement loss $\mathcal{L}_{D}$ and the class diversity loss $\mathcal{L}_{div}$, with a weighting factor of $\lambda$.

\begin{equation}
    \mathcal{L}_{G}=\mathcal{L}_{D}+\lambda \mathcal{L}_{div}
\end{equation}
\subsubsection{Defender goal}
In regards to model cloning attacks, the defender has to stop an adversary from being able to clone the victim model. The objective of a defender is to reduce the test accuracy of the cloned model while maintain high accuracy for benign users of the service. The constraints can be expressed by setting a threshold, T, for model accuracy on in-distribution cases.  The Defender goal can be formulate to minimize the accuracy of cloning model $Acc[\mathcal{M'}_{v}(x; \theta')]$ on the victim's target distribution $\mathcal{D}(X)$ with threshold T.

\begin{equation}
\label{mis_eq2}
    \min_{\theta'} \mathbb{E}_{x \sim \mathcal{D}(X)} [\mathcal{M'}_{v}(x;\theta')]
\end{equation}

\begin{equation}
    \mathbb{E}_{x \sim \mathcal{D}(X)} [\mathcal{M'}_{v}(x; \theta')] \le T
 \label{eq:T}
\end{equation} 

Equations \ref{mis_eq2} and \ref{eq:T} present a constrained optimization problem for the defender, enabling a balance between model accuracy and security against model extraction attacks. Defending within these constraints, such as our suggested defense, is called accuracy-constrained.These techniques offer enhanced security, without compromising classification accuracy.

\section{Proposed Methodology}
In this section, proposed countermeasure MisGUIDE framework to
defend the model extraction attacks is described. Subsection 3.1 provides a brief overview to proposed MisGUIDE framework, while Subsections 3.2 and 3.3 describe the two main components of MisGUIDE: the vision transformer acting as an  OOD detector and the probabilistic threshold criteria. 

\subsection{MisGUIDE Framework}

The MisGUIDE defense mechanism relies on the insight that contemporary 
model extraction attacks leverage a generative framework to create new query samples from a distribution using random noise, notably generating numerous OOD samples for victim model queries. The key principle of MisGUIDE is to employ a Vision Transformer as an OOD detector to detect potentially malicious query from adversaries. In accordance with the proposed probabilistic threshold, MisGUIDE deliberatly furnishes inaccurate predictions for queries identified as OOD, preserving accurate predictions for in-distribution (ID) queries. This strategy leads to the mislabeling of a substantial portion of the adversary's dataset. Training a model on this mislabeled dataset yields a clone model with poor generalization accuracy.
 
Figure \ref{fig:Model} provides a visual depiction of proposed MisGUIDE defense framework. The defense comprises four integral components: (1) Victim Model $M$, (2) an OOD detector, (3) a misguiding probabilistic threshold, and (4) a response switching mechanism. The switch mechanism determines whether the system should respond with a correct or incorrect prediction to the input query $x \in \mathcal{R}^d$ based on the output of the proposed probabilistic misguiding threshold $p$. This schematic representation emphasizes the essential components of our defense strategy, underscoring 
its versatility and multifaceted approach in mitigating adversarial scenarios.

Algorithm \ref{alg:MisGUIDE}, details the key steps and working flow of the MisGUIDE framework.
In Algorithm \ref{alg:MisGUIDE}, $x$ denotes a $d$-dimensional query sample vector in $\mathcal{R}^d$, $x_{input}^{Emd} \in \mathcal{R}^e$ denotes $e$-dimensional embedding vector for query $x$ computed using 
ViT, $IS_{ood} \in \lbrace True, False \rbrace$ denotes a boolean flag for OOD, $r \in \mathcal{N}(0,1)$ denotes a random number, $p$ denotes a probabilistic threshold hyperparameter fixed by the user, and $\mathcal{M}$ denotes the victim model.

While processing an input query $x$, MisGUIDE initiates by assessing whether the input is OOD or not. If the input query is determined to be within the victim model's learned distribution, MisGUIDE treats it as originated from a benign user, and therefore, accurate predictions are provided for the input query. Conversely, if query $x$ is identified as an OOD input query, indicating potential malicious intent, MisGUIDE employs a switch mechanism unlike the other defense mechanisms that uniformly provide inaccurate predictions for all OOD input queries. This switching mechanism selectively determines whether to offer accurate or inaccurate predictions based on the functionality of the misguiding function. This enables MisGUIDE to randomly respond to OOD input queries with a probabilistic threshold, providing a nuanced and adaptable defense strategy.

\begin{algorithm}[H]
\caption{MisGUIDE}
\label{alg:MisGUIDE}
\begin{algorithmic}[1]
\Require Input query $x \in \mathcal{R}^d$
\Ensure Prediction logit score vector: $Pred$
    \State $x_{input}^{Emd}\gets\text{ViT}(x)$ \qquad \text{//}~~$x_{input}^{Emd} \in \mathcal{R}^e$
    \State $IS_{ood}\gets\text{OOD}(x_{input}^{Emd})$ \qquad \text{//}~~$IS_{ood} \in \lbrace True, False \rbrace$
    
    \If{$IS_{ood} = \text{True } $}
        \State $r \gets \text{random number}\in \mathcal{N}(0,1)$
        \If{$(r \le p)$}
            \State $Pred \gets \text{Random Logit}$
        \Else
        \State $Pred \gets \mathcal{M}(x)$
        \EndIf
    \Else
        \State $Pred \gets \mathcal{M}(x)$
        
    \EndIf
 \State \Return $Pred$
\end{algorithmic}
\end{algorithm}

{\bf Time Complexity of MisGUIDE:} Let's assume the time complexity of the ViT model is $O(V)$ for a query vector $x$, the time complexity for detecting OOD for embedding vector $x_{input}^{Emd}$ is $O(D)$ and the time complexity for generating inference for a query $x$ from victim model $\mathcal(M)$ is $O(I)$.  The total time complexity of  Algorithm \ref{alg:MisGUIDE} is $O(V + D + I)$.

The subsequent section delves into a more comprehensive explanation of the OOD detector using ViT and Misguiding probabilistic threshold.

\subsection{Vision Transformer: an OOD detector}

This section presents pre-trained ViT model as an OOD detector for safeguarding the victim model against model extraction attacks. 
The Vision Transformer (ViT), proposed by Dosovitskiy et al. in 2021 \cite{dosovitskiy2021vit}, leverages self-attention modules 
and extensive pre-training, demonstrating efficacy in various vision tasks. Notably, ViT exhibits performance on par with other models, coupled with efficient training and quick adaptability to smaller datasets, rendering it a compelling choice as a backbone for tasks such as  OOD detection. Our proposed OOD detection algorithm utilizes pre-trained ViT model trained on the ImageNet-21k dataset \cite{ridnik2021imagenet21k}.

\begin{algorithm}[H]
\caption{OOD Training}
\label{ood_train}
\begin{algorithmic}[1]
\Require Training samples $\{x_i^{in}\}$ (in-distribution), training labels $\{y_i^{in}\}$ for $i = 1, \ldots, n$ and number of classes $C$
\Ensure OOD parameters $\mu$ and $\Sigma$ for all classes for Victim Model's In-data Distribution.

\For{each class $c = 1$ to $C$}
    \State  $\{x_i^{\text{Emd}}\} = ViT_{\text{}}(\{x_i^{\text{in}} | y_i^{\text{in}} = c\})$
    \State $\mu_c = \frac{1}{n_c} \sum_{i=1}^{n_c} x_i^{\text{Emd}}$
    \State $\Sigma_c = \frac{1}{n_c - 1} \sum_{i=1}^{n_c} (x_i^{\text{Emd}} - \mu_c)(x_i^{\text{Emd}} - \mu_c)^T$
\EndFor \\
\Return $\mu = [\mu_1, \mu_2, \ldots, \mu_C ] $, $\Sigma = [\Sigma_1, \Sigma_2, \ldots, \Sigma_C]$
\end{algorithmic}
\end{algorithm}

\begin{algorithm}[H]
\caption{OOD Inference}
\label{ood_infer}
\begin{algorithmic}[1]
\Require Query sample $x_{\text{test}}$
\Ensure Query $x_{\text{test}}$ is OOD or not (a boolean flag (True/False)
\State  $x_{\text{test}}^{\text{feat}} = ViT_{\text{}}(x_{\text{test}})$
\State  $\text{distance} = \min_c \left[ (x_{\text{test}}^{\text{feat}} - \mu_c)^T \Sigma_c^{-1} (x_{\text{test}}^{\text{feat}} - \mu_c) \right]$

\If{$\text{distance} > t_{\text{distance}}$ }
    \State flag = True \quad // $x_{\text{test}}$ is OOD
\Else~
    \State flag =False \quad // $x_{\text{test}}$ is not in OOD
\EndIf
\Return flag
\end{algorithmic}
\end{algorithm}

The OOD training Algorithm \ref{ood_train} learns the distribution of the training data used by victim model during its training.
We explore a scenario in which the model undergoes fine-tuning exclusively on the in-distribution data (training set $\mathcal{X}_{t}$) without any knowledge of the OOD data. The OOD training algorithm is designed for OOD detection using Visual Transformer (ViT) embeddings. It begins by taking as input a set of training samples with $C$ classes ( $x_{i}^{in} \in \mathcal{X}_{t}$ ) and their associated labels ($y_{i}^{in}$). The primary objective is to determine whether the input query sample is an OOD or not. OOD training algorithm checks the belongingness of input query $x$ with respect to known class $c$. The algorithm first computes ViT embeddings for each training class by applying the ViT model to the respective training samples $x_{i}^{in}$. Subsequently, it calculates and returns the mean ($\mu$) and covariance matrix ($\Sigma$) of these embeddings for each class, serving as statistical descriptors of the distribution. 

{\bf Time Complexity of OOD Training:} The time complexity for step 1 and step 5 of Algorithm \ref{ood_train} is $O(\sum_c^C (n_c e V)$. The total time complexity of Algorithm \ref{ood_train} can be computed by $O(ne V)$.

The OOD inference Algorithm \ref{ood_infer} outlines an OOD detection method for a query sample $x_\text{test}$.
This algorithm starts by computing $e$-dimensional embedding vector for the test query via ViT, and the Mahalanobis distance \cite{lee2018simple} is computed using this test query embedding vector and the in-distribution parameters (Algorithm \ref{ood_train})  mean $\mu$ and covariance matrix $\Sigma$ using Equation \eqref{dis}. 
\begin{equation}
\label{dis}
\text{distance} = \min_c \left[ (x_{\text{test}}^{\text{Emd}} - \mu_c)^T \Sigma_c^{-1} (x_{\text{test}}^{\text{Emd}} - \mu_c) \right]
\end{equation}

The minimum distance across all classes is selected, and the test query is classified as an OOD if the Mahalanobis distance is greater than a predefined threshold ($tdistance$). Conversely, if these conditions are not met, the test sample is considered as in-distribution data.

{\bf Time Complexity of OOD Inference:} The time complexity for step 1 
 and step 2 of Algorithm \ref{ood_infer} is $O(V)$  and $O(Ce^3)$. The total time complexity of Algorithm \ref{ood_infer} can be computed by $O(Ce^3 + V)$.

\subsection{Misguiding probabilistic Threshold $p$}
In the proposed defense mechanism, we learn a OOD detector using the vision transformer and probabilistically preventing the model to generate a random output for OOD samples which will limit the expressibility of the attacking the model. In the proposed MisGUIDE Algorithm \ref{alg:MisGUIDE}, step 5 introduces a probabilistic threshold denoted as
$p$ to regulate the random predictions made by the victim model. When 
$p$ is set to a high value, a large proportion of queries receive random labels; conversely, a low $p$ results in a significant number of queries obtaining correct labels from the victim model. Specifically, for 
$p$= 0, all queries receive correct labels from the victim model, while for $p$=1, all queries are assigned random labels. This characteristic 
makes MisGUIDE a versatile framework, offering a spectrum of random responses for  OOD queries.

\section{Experiments}
In this section, we evaluate the efficacy of the proposed defense mechanism and compare it with state of art defense approaches with comprehensive experiments. Standard experiment setup is described in Section 4.1, and the defensive abilities of different approaches against various attack methods are investigated in Section 4.2. Furthermore, the research examines a few drawbacks of the misguidance method in Section 4.3.
\subsection{Experiment Setup}
This research evaluates the defensive performance against the following black-box model cloning attack methods:

\begin{itemize}
    \item DFME\cite{Truong2021}: DFME attacks are also a form of data-free model extraction method that does not necessitate a surrogate dataset. This approach incorporates techniques from the field of data-free knowledge transfer for model extraction. DFME has identified that the selection of the loss function between student-teacher predictions is crucial to ensuring that the extracted model accurately replicates the victim model. 
    
    \item DisGuide (Soft label and Hard label) \cite{Rosenthal2023}: DisGuide is also a data-free model extraction method that utilizes data-generated input queries through a generative model. This approach is capable of achieving higher accuracy in terms of cloning the victim model and requires a lower number of queries compared to DFME and MAZE. This method operates effectively under both soft label and hard label attack settings.    
\end{itemize}

The proposed MisGUIDE method is compared with the following defence methods:
\begin{itemize}
\item \textbf{Without Defense}:A non defensive mechanism is employed with the victim model, ensuring that the original prediction is consistently delivered to the user without any perturbation.

\item \textbf{Adaptive Misinformation (AM)}\cite{kariyappa2020defending}  
is a defense technique aimed at preventing model cloning attacks. It works by identifying that cloning attacks typically use OOD inputs. The defense strategically alters predictions for every queries adeptly, effectively lowering the accuracy of the adversary's copied model while preserving the accuracy of authenticate users. 
\item MisGUIDE: Our Proposed Defense Framework.
\end{itemize}

\textbf{Datasets, Victim, and Clone Architectures:} In our experiments, we utilize two commonly employed image classification datasets, particularly CIFAR-10 (Krizhevsky, Hinton et al. 2009), in alignment with the methodology of prior studies such as \cite{Truong2021} \cite{Sanyal2022, Rosenthal2023, kariyappa2020defending}. The focus of our evaluation is on assessing the performance of the proposed defense mechanism MisGUIDE against the DFME and DisGuide attacks. To evaluate the effectiveness of MisGUIDE against these attacks, we follow the same settings of the parameters as outlined in \cite{Truong2021, Rosenthal2023}. These attacks involve extracting the functionality of a ResNet-34 victim model and transplanting it into ResNet-18 clone models. The ResNet-34 victim models employed in our experiments exhibit a test accuracy of 95.54\% on the CIFAR-10 dataset and test accuracy of 77.52\% on CIFAR-100.

\textbf{Attack Hyperparameters and Settings} 

The proposed MisGUIDE framework is embedded with  Opensource implementation of Disguide and DFME \footnote{Disguide and DFME \url{https://github.com/lin-tan/disguide}}.  We followed the same attack settings proposed in \cite{Truong2021, Rosenthal2023}, except for the training batch size. A training batch size of $2048$ is used to train the clone model. 

\textbf{OOD Settings}: We utilize the Vision Transformer (ViT) architecture by \cite{dosovitskiy2021vit} with pretrained model checkpoints from ImageNet-21. Fine-tuning involves training the full ViT on CIFAR-10 or CIFAR-100, and we abstain from using data augmentation to enhance generalization. Pre-logit embeddings of size 1024 are extracted from the fine-tuned ViT for both CIFAR-10 and CIFAR-100, serving as representations for OOD tasks. Mahalanobis distance using Equation \eqref{dis} is employed for OOD detection, measuring dissimilarity between embedding and enabling robust identification of instances deviating from the learned distribution.

\subsection{Results}
We assess the effectiveness of the MisGUIDE defense strategy against two state-of-the-art data-free model extraction methods. In the soft-label setting, we compare MisGUIDE with both DisGUIDE and DFME. However, in the hard-label setting, our comparison is limited to DisGUIDE alone, as the DFME paper did not provide evaluations for DFME in the hard-label setting. Additionally, we assess the OOD detector's performance by comparing it to the Maximum over Softmax Probabilities (MSP) method. This evaluation aims to gauge the effectiveness of the OOD detector in distinguishing OOD instances, with MSP serving as a benchmark for comparison.

\subsubsection{OOD detecter performance}
\begin{figure}
  \centering
   \includegraphics[width=1\linewidth]{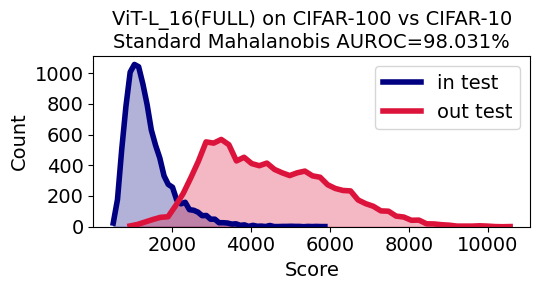}
   \caption{Distribution plot for CIFAR-10 and CIFAR-100}
   \label{fig_dis_plot}
\end{figure}

\begin{figure}
  \centering
   \includegraphics[width=1\linewidth]{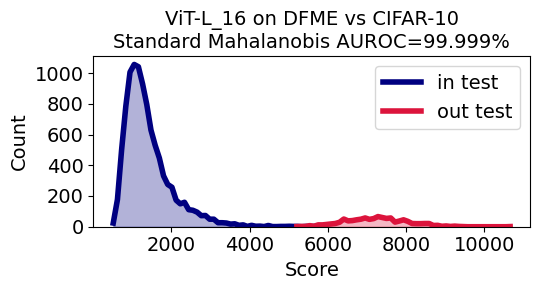}
   \caption{Distribution plot for CIFAR-10 and DFME queries}
   \label{fig_dis_dfme_plot}
\end{figure}

The table \ref{tab:vit_matrics} presents a comparative analysis of the Vision Transformer (ViT) model's performance in various scenarios, specifically focusing on in-distribution and OOD datasets. 
We have calculatea the Area Under the Receiver Operating Characteristic Curve (AUROC) derived from prediction scores produced by our OOD Detector implementation, treating the task as a binary classification problem distinguishing between In-Distribution (ID) and OOD samples. Figure \ref{fig_dis_plot} and \ref{fig_dis_dfme_plot} describes the distribution of CIFAR-10 vs CIFAR-100 and CIFAR-10 vs DFME queries dataset.

For in-distribution datasets, the ViT model achieves an accuracy of 98.10\% on CIFAR-10 and 93.26\% on CIFAR-100. After fine-tuning, the model demonstrates improved test accuracies of 98.42\% on CIFAR-10 and 95.53\% on CIFAR-100. The OOD evaluation involves datasets corresponding to the opposite CIFAR variant (CIFAR-100 for CIFAR-10 and vice versa). In terms of OOD detection, the ViT model exhibits robust performance as indicated by Mahalanobis AUROC scores of 97.68\% for CIFAR-100 and 95.53\% for CIFAR-10. Additionally, the Maximum over Softmax Probabilities (MSP) method is employed for comparison, yielding AUROC scores of 97.68\% for CIFAR-100 and 91.89\% for CIFAR-10. In Figure \ref{fig_tsne_2D} and Figure \ref{fig_tsne_3D}, illustrates the data distribution using t-SNE for ViT embeddings for CIFAR-10, CIFAR-100 and GAN generated quesries. It can be observed from these figures that the distribution varies across the datasets. This plays a key role in detecting the threshold for OOD queires in our proposed MisGUIDE defense mechanism.

\begin{table*}[]
\centering
\begin{tabular}{l l l l l l}
\hline
\textbf{Model ID} & \textbf{ID} & \textbf{FT-Accuracy} & \textbf{OD-Dataset} & \textbf{Mahalanobis-AUROC} & \textbf{MSP-AUROC} \\
\hline
ViT & CIFAR-10 & 98.10\% & CIFAR-100 & 98.03\% & 97.68\% \\
ViT & CIFAR-100 & 93.26\% & CIFAR-10 & 96.21\% & 92.99\% \\
\hline
\end{tabular}
\caption{OOD detector performance metrics for Cifar 10 and Cifar 100 fine tuned as In-distribution and tested with Cifar100 and Cifar10 out distribution.}
\label{tab:vit_matrics}
\end{table*}

\begin{table*}[h]
\centering
\begin{tabular}{@{}lcccccc@{}}
\toprule
\textbf{Attack Method} & \textbf{Settings} & \textbf{Victim (\%)} & \textbf{None (\%)} & \textbf{AM (\%)} & \textbf{MisGUIDE (\%)} \\
\midrule
DFME & Soft-label & 95.5 & 88.4 & 14.0 & 22.5 \\
DisGuide & Soft-label & 95.5 & 94.2 & 13.4 & 24.3 \\
Disguide & Hard-Label & 95.5 & 87.5 & 22.6 & 27.3 \\
\midrule
\textbf{Test Accuracy} & - & - & 95.5 & 77.2 & 91.8 \\
\bottomrule
\end{tabular}
\caption{Cloning accuracy (\%) for various Defense Strategies for CIFAR-10 dataset}
\label{tab:Mis_guide}
\end{table*}

\begin{figure*}[!ht]
    \centering
\begin{subfigure}[b]{0.45\textwidth}
   \includegraphics[width=\textwidth]{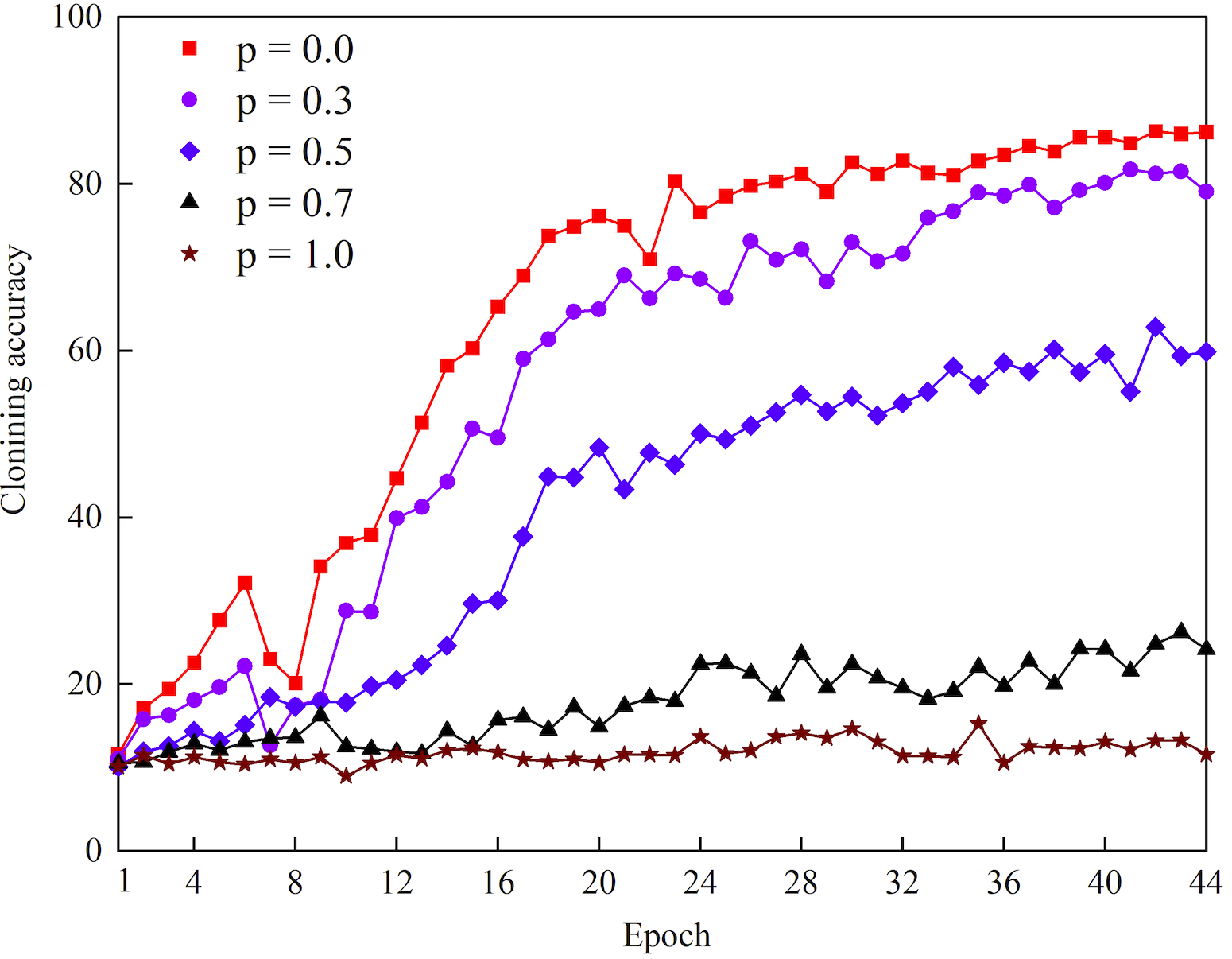}
   \caption{CA of MisGUIDE for DFME attack}
   \label{fig:dfme_p}
\end{subfigure}
\begin{subfigure}[b]{0.45\textwidth}
   \includegraphics[width=\textwidth]{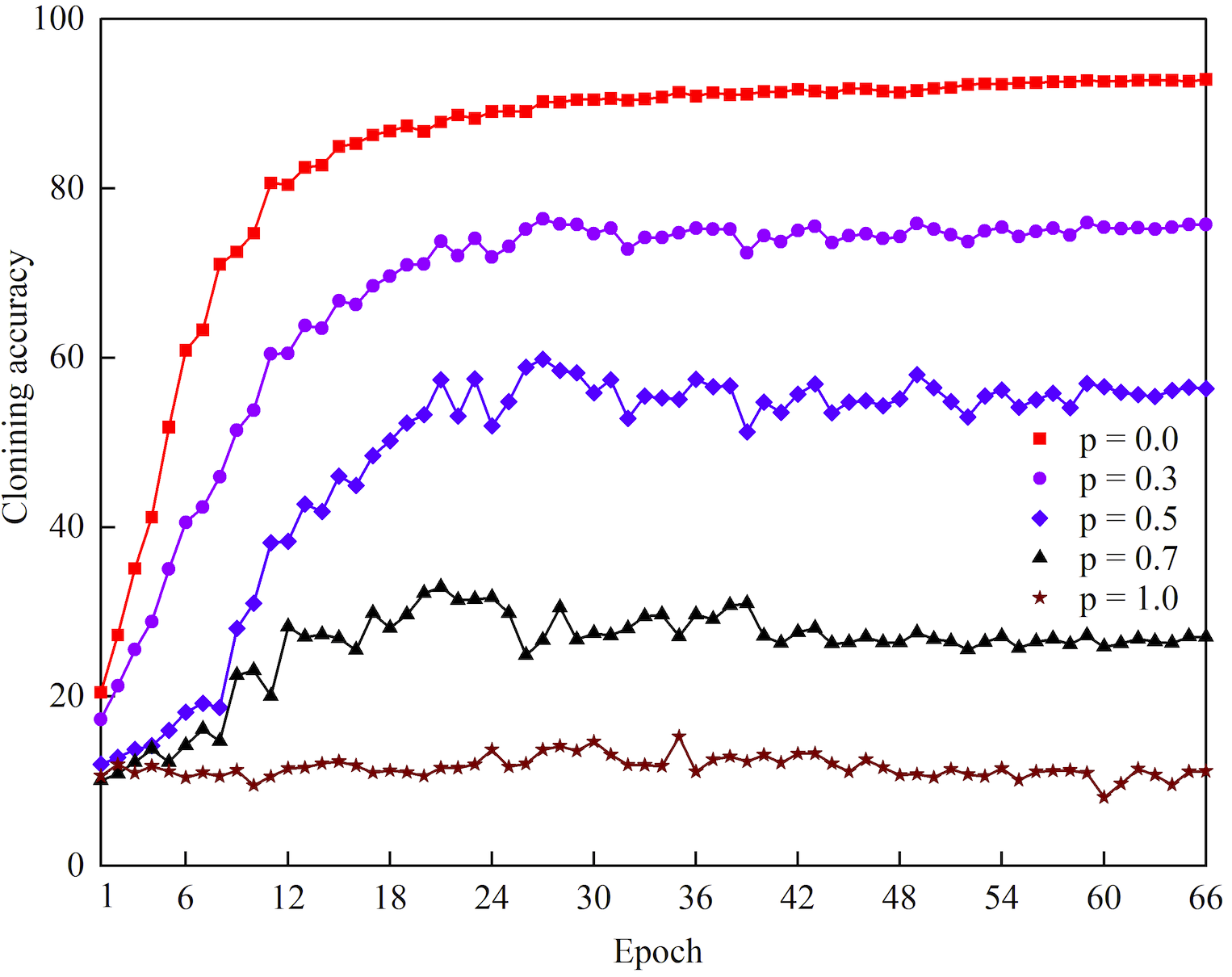}
   \caption{CA of MisGUIDE for DisGUIDE (SL) attack}
   \label{fig:dgsl_p}
\end{subfigure}
\medskip
\begin{subfigure}[b]{0.45\textwidth}
   \includegraphics[width=\textwidth]{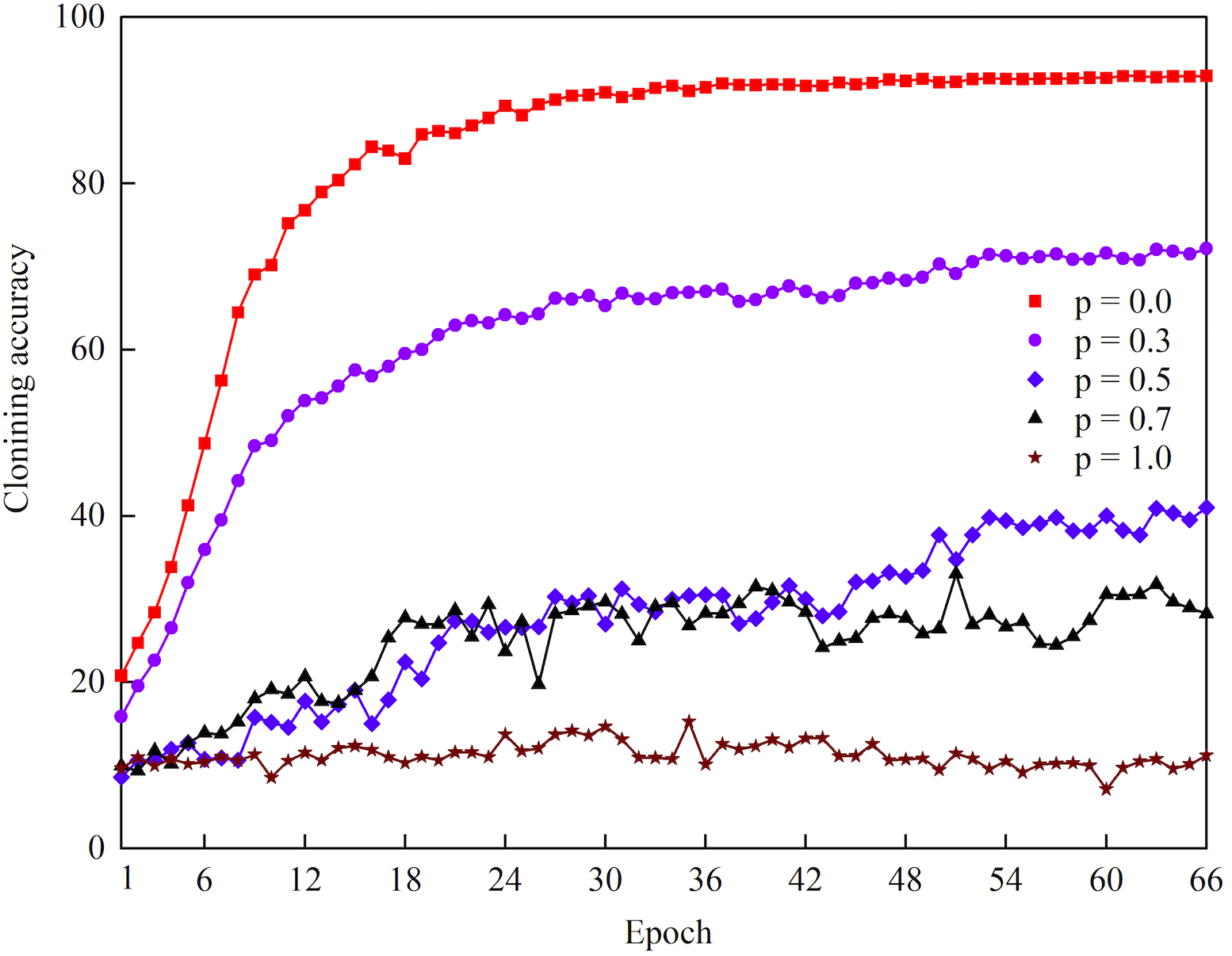}
   \caption{CA of MisGUIDE for DisGuide (HL) attack}
   \label{fig:dghl_p}
\end{subfigure}
\begin{subfigure}[b]{0.45\textwidth}
    \includegraphics[width=\textwidth]{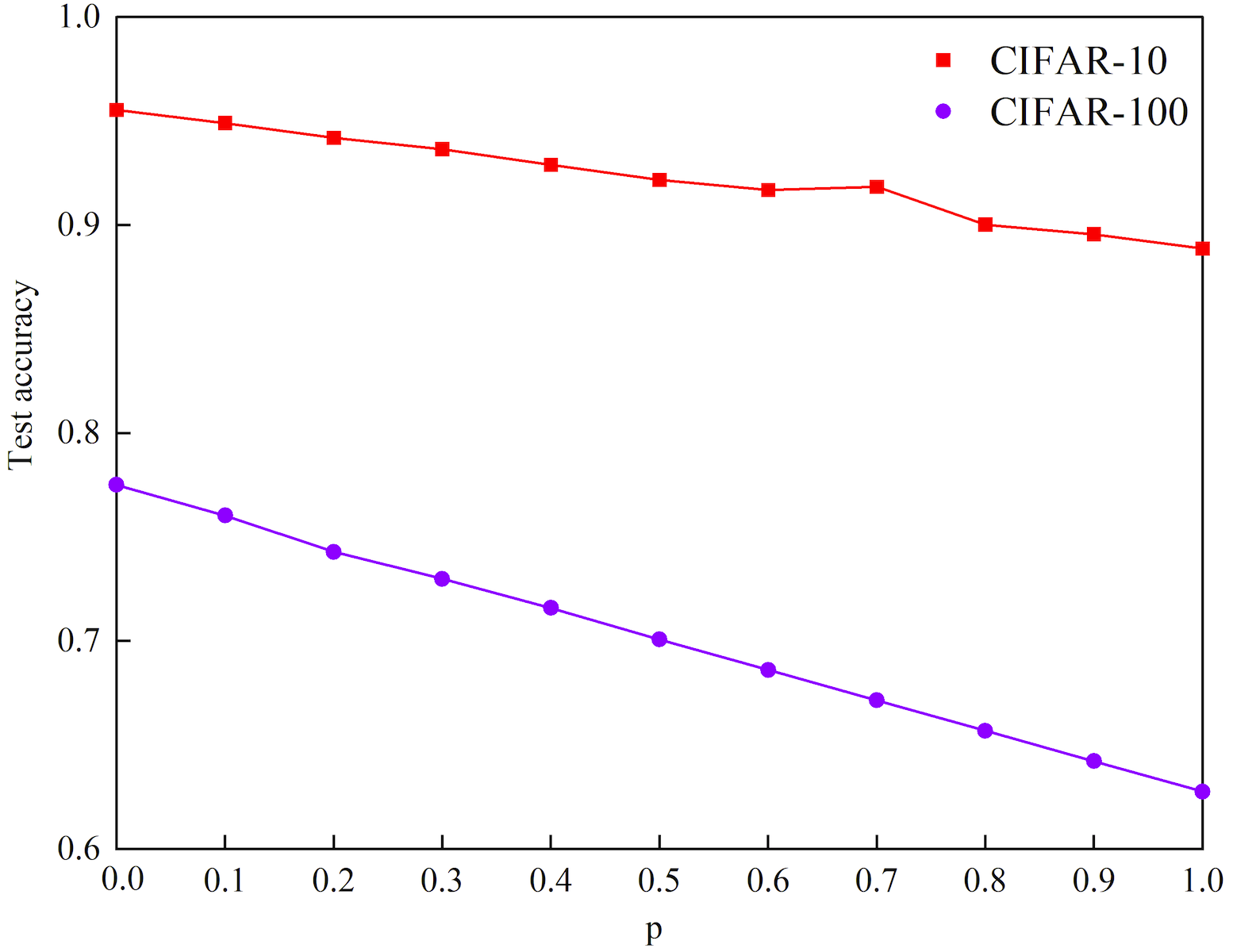} 
    \caption{Accuracy of Victim model on different values of $p$.}
    \label{fig_vary_p}
\end{subfigure}
    \caption{Assessment of Cloning Accuracy (CA) for varying Probabilistic Threshold $p$}
    \label{fig_prob}
\end{figure*}

\begin{figure}
  \centering
   \includegraphics[width=1\linewidth]{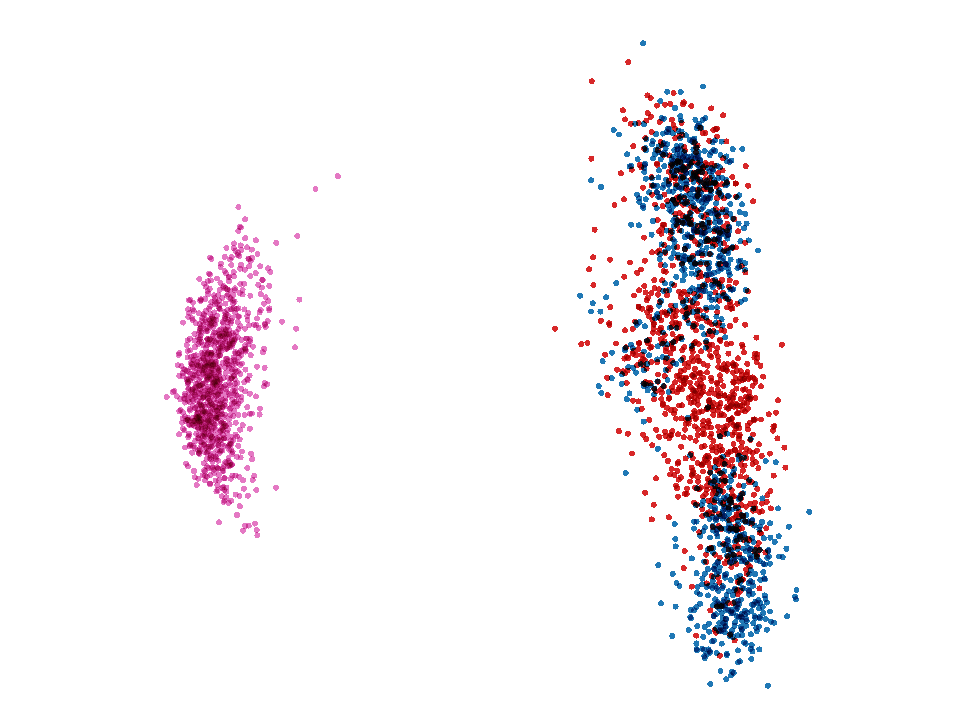}
   \caption{Plot of embeddings in 2D using t-SNE for CIFAR-10, CIFAR-100 and GAN Generated Queries}
   \label{fig_tsne_2D}
\end{figure}

\begin{figure}
  \centering
   \includegraphics[width=1\linewidth]{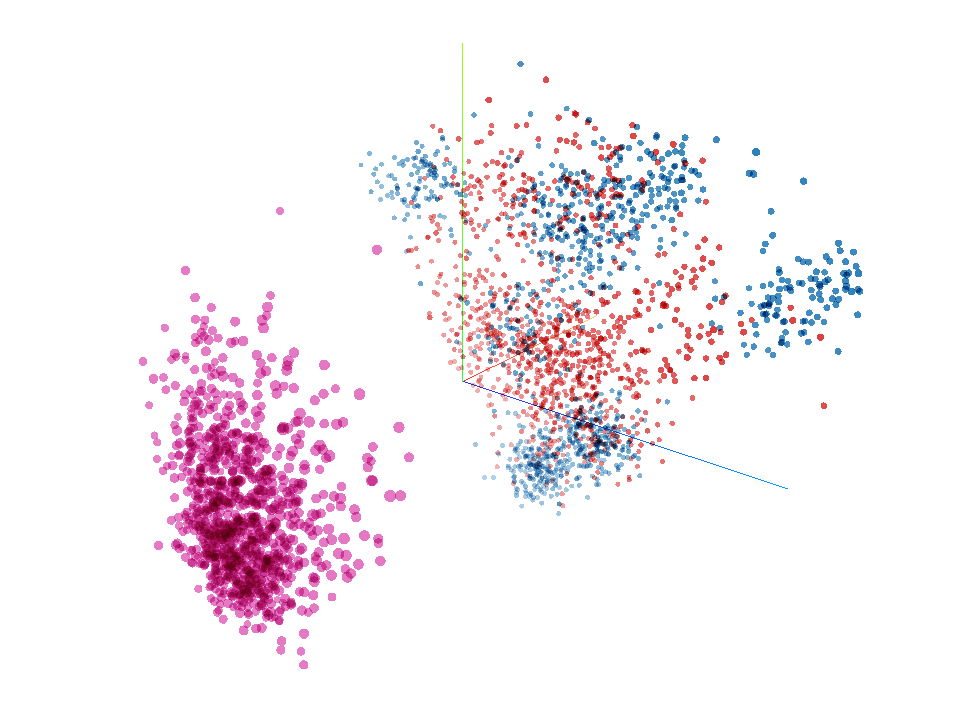}
   \caption{Plot of embeddings in 3D using t-SNE for CIFAR-10, CIFAR-100 and GAN Generated Queries}
   \label{fig_tsne_2D}
\end{figure}

It can be observed from the Figure \ref{fig:dfme_p}, \ref{fig:dgsl_p}, \ref{fig:dghl_p} and \ref{fig_vary_p} that  the value $p$ = 0.7 seems to represent a significant point where the trade-off is optimized clone accuracy is substantially reduced, indicating less effective cloning, without a corresponding decrease in test accuracy, which remains high. This suggests that $p$ = 0.7 is a threshold at which the model maintains its robustness against cloning attempts while still performing well on the victim test. This balance is critical in scenarios where both high performance and security (or uniqueness) are valued.

\subsubsection{Cloning Accuracy}
The table \ref{tab:Mis_guide} presents a comprehensive comparison of final clone accuracy under different conditions, particularly focusing on attacks DFME and DisGUIDE, along with the application of the MisGUIDE defense. For each attack scenario, the table delineates the final clone accuracy both in the absence of any defense mechanism and when defended using MisGUIDE. The statistical results reveal significant improvements with MisGUIDE, evidenced by the substantial reduction in final clone accuracy for both DFME and DisGUIDE attacks. Without defense, DFME achieved a final clone accuracy of 88.4\%, whereas with MisGUIDE, this accuracy decreased to 22.5\%. Similarly, for DisGUIDE, MisGUIDE demonstrated a reduction from 94.2\% to 24.3\% in clone accuracy in soft label setting and reduction from 87.5\% to 27.3\% in clone accuracy when countered by MisGUIDE in hard label setting for $p$ = 0.7. Further, the MisGUIDE achieves a better defense against AM method with high test accuracy. The similar results are obtained for both CIFAR-10 and CIFAR-100 dataset. 

In order to assess the efficacy of our proposed defense mechanism, MisGUIDE (MG), we subjected it to evaluation against a range of adversarial attacks, including the Naive Attack, Top-1 Attack, S4L Attack, and pBayes Attack using KnockoffNet query method \cite{tang2024modelguard}. The resulting cloning accuracies for these attacks against MG defense method have been  observed as 15.17\%, 13.63\%, 17.48\%, and 16.23\%, respectively. Notably, throughout these attacks, the protected accuracy of our defense mechanism remained consistently high at 93.78\%. It is noteworthy that even our least effective defense against cloning outperformed the current state-of-the-art defense model, ModelGuard, which achieved cloning accuracies of 75.06\%, 83.51\%, 71.23\%, and 85.63\% under similar attack conditions.
These experimental insights highlight the efficacy of MisGUIDE in enhancing the resilience of the cloned model against adversarial extraction attempts, thereby reinforcing the model's security and robustness.

\textbf{Impact of parameter $p$ on cloning accuracy}
The effect of cloning accuracy with varying $p \in [0, 1]$ for CIFAR10 dataset is illustrated in Figure \ref{fig_prob} (\ref{fig:dfme_p},  \ref{fig:dgsl_p} and \ref{fig:dghl_p}) for DFME, DisGUIDE with Soft Label and DisGUIDE with Hard Label respectively. From these figures, it can be observed that for low value of the probabilistic threshold $p$ (less defense), the cloning accuracy is high while for the high value of $p$ (hard defense), the cloning accuracy is very low. From these figures and Table \ref{tab:Mis_guide}, it can be observed that $p$ = 0.7 can be chosen a trade-off between the cloning accuracy and the test accuracy.

\section{Conclusion}
The escalating threat of model cloning attacks presents significant challenges to the privacy and intellectual property of machine learning models. Existing defensive approaches face obstacles such as increased computational overhead and undesirable trade-offs, limiting their practicality and effectiveness, particularly in hard label scenarios. Addressing this issue, this paper introduces an innovative and versatile 
defense framework, MisGUIDE.

This framework integrates countermeasures specifically designed to 
mislead data-free model attacks in a probabilistically controlled manner, 
applicable to both hard label and soft label settings. The distinguishing 
feature lies in the OOD detection module and probabilistic threshold-based control, setting it apart from similar existing defenses such 
Adaptive Misinformation \cite{Kariyappa2021}. Comprehensive 
experiments underscore the effectiveness and practicality of MisGUIDE as a solution to fortify the integrity and security of machine learning models. Future work will focus on establishing the suitability of the MisGUIDE framework in comparison to the recently proposed defense 
ModelGuard \cite{tang2024modelguard}.

{\footnotesize \bibliographystyle{acm}
\bibliography{sample}}

\theendnotes

\end{document}